\documentclass[preprintnumbers,amsmath,amssymb,showpacs]{revtex4}

\usepackage{graphicx}
\usepackage{dcolumn}
\usepackage{bm}

\begin{document}

\title{Separability criteria for several classes of  $n$-partite quantum states}

\author{Ting Gao}
\email{gaoting@hebtu.edu.cn}
\author{Yan Hong}
\affiliation {College of Mathematics and Information Science, Hebei
Normal University, Shijiazhuang 050016, China}

\date{\today}

\begin{abstract}
In this paper, we mainly discuss the separability of $n$-partite
 quantum states from elements of density matrices. Practical separability criteria for different classes of $n$-qubit
and $n$-qudit  quantum states  are obtained. Some of them are also
sufficient conditions for genuine entanglement of $n$-partite
quantum states.  Moreover, one of the resulting criteria is also
necessary and sufficient for a class of $n$-partite states.
\end{abstract}


\maketitle

\section{Introduction}

Quantum entanglement is a kind of new resources beyond the classical
resources, and has widely been applied to quantum communication
\cite{Ekert91, BBCJPWteleportation93, BBM92, LoChau, YZEPJB05,
GYJPA05} and quantum computation \cite{RB,BD}. Whether a state is
entangled or not is one of the most challenging open problems. For
the states of $2\times 2$ and $2\times 3$ bipartite systems, they
are separable iff they are positive partial transposition (PPT)
\cite{PeresPPT, HorodeckiPPT}. For high dimensional and multipartite
systems, however, the situation is significantly more complicated,
as several inequivalent classes of multiparticle entanglement exist
and it is difficult to decide to which class a given state belongs.

It would be desirable to have useful criteria that allow us to
detect the different classes of multipartite entanglement directly
from a given density matrix. G\"{u}hne and Seevinck
\cite{GuhneNJP2010} presented a method to derive separability
criteria for different classes of 3-qubit and 4-qubit entanglement,
especially genuine 3-qubit and 4-qubit entanglement. Huber et al.
\cite{MarcusPRL2010} developed a general framework to identify
genuinely multipartite entangled mixed quantum states in
arbitrary-dimensional systems. Based on the framework,
k-separability criterion was derived in \cite{ABMarcus1002.2953}.

In this paper, the separability of $n$-partite and multilevel
quantum states from elements of density matrices is investigated. We
derive simple algebraic tests, which are necessary conditions for
separability of  $n$-partite  quantum states. Some of them are also
sufficient conditions for   genuine entanglement of $n$-qubit and
$n$-qudit quantum states. One of the resulting criteria is necessary
and sufficient for a certain family of $n$-partite states.

An $n$-partite pure state $|\psi\rangle\in \mathcal{H}_1\otimes
\mathcal{H}_2\otimes\cdots \mathcal{H}_n$ is called biseparable if
there is a bipartition $j_1j_2\cdots j_k|j_{k+1}\cdots j_n$ such
that
\begin{equation}\label{}
|\psi\rangle=|\psi_1\rangle_{j_1j_2\cdots
j_k}|\psi_2\rangle_{j_{k+1}\cdots j_n},
\end{equation}
where $|\psi_1\rangle_{j_1j_2\cdots j_k}$ is the state of particles
$j_1,j_2,\cdots, j_k$, $|\psi_2\rangle_{j_{k+1}\cdots j_n}$ is the
state of particles $j_{k+1},\cdots, j_n$, and  $\{j_1,j_2,\cdots,
j_n \}=\{1,2,\cdots, n\}$. An $n$-partite mixed state $\rho$ is
biseparable if it can be written as a convex combination of
biseparable pure states
\begin{equation}\label{}
 \rho=\sum\limits_{i}p_i|\psi_i\rangle \langle\psi_i|,
\end{equation}
where $|\psi_i\rangle$ might be biseparable under different
partitions.  If an $n$-partite state is not biseparable, then it is
called genuinely $n$-partite entangled. Genuine $n$-partite
entanglement  is very important as one usually aims to generate and
verify this class of entanglement in experiments
\cite{GuhnePhysRep09}. We mainly discuss entanglement criteria for
this type of entanglement.
 An $n$-partite pure state is  fully separable if it is of the
form
\begin{equation}\label{}
|\psi\rangle=|\psi\rangle_1|\psi\rangle_2\cdots|\psi\rangle_n,
\end{equation}
and an $n$-partite mixed state is fully separable if it is a mixture
of fully separable pure states
\begin{equation}\label{}
 \rho=\sum_i p_i |\psi_i\rangle \langle\psi_i|,
\end{equation}
where the $p_i$ forms  a probability distribution, and
$|\psi_i\rangle$ is fully separable.
  We also consider
separability criteria of biseparable and fully separable $n$-qubit
and $n$-qudit states, and give clear and complete proof of each
criterion from general partition by using the Cauchy inequality and
H\"{o}lder inequality.

\section{The separability criteria of biseparable  $n$-partite  states and genuine $n$-partite entangled states }

Let $\rho$ be a density matrix describing an $n$-particle system,
whose state space is Hilbert space $\mathcal{H}_1\otimes
\mathcal{H}_2\otimes\cdots \mathcal{H}_n$, where
dim$\mathcal{H}_l=d_l$, $l=1,2,\cdots, n$. We denote its entries by
$\rho_{i,j}$, where $1\leq i,j\leq d_1d_2\cdots d_n$.

Next we investigate biseparable  $n$-partite  states and genuine
$n$-partite entangled states.

\textbf{Theorem 1} (G\"{u}hne and Seevinck \cite{GuhneNJP2010}) For
any $n$-qubit density matrix, $\rho=(\rho_{i,j})_{2^n\times 2^n}$,
if it is biseparable, then
\begin{equation}\label{n-qubit-biseparable}
    |\rho_{1,2^n}|\leq\sum_{i=2}^{2^{n-1}}\sqrt{\rho_{i,i}\rho_{2^n-i+1,2^n-i+1}}=\frac{1}{2}\sum_{i=2}^{2^n-1}\sqrt{\rho_{i,i}\rho_{2^n-i+1,2^n-i+1}}.
\end{equation}
That is, if the inequality (\ref{n-qubit-biseparable}) does not
hold, then $\rho$ is a genuine $n$-qubit entangled state.

\textbf{Proof.} First we show that (\ref{n-qubit-biseparable}) holds
for pure state.

Suppose that $\rho=|\psi\rangle\langle\psi|$ is an $n$-qubit pure
biseparable state under the $j_1j_2\cdots j_k|j_{k+1}\cdots j_n$
partition, and
\begin{equation}\label{}
 \begin{array}{rl}
   |\psi\rangle= & |\phi_1\rangle_{j_1j_2\cdots j_k}|\phi_2\rangle_{j_{k+1}\cdots j_n} \\
    = & (\sum\limits^1_{i_1,i_2,\cdots, i_k=0}a_{i_1i_2\cdots i_k}|i_1i_2\cdots
    i_k\rangle)_{j_1j_2\cdots j_k}
   (\sum\limits^1_{i_{k+1},\cdots, i_n=0}b_{i_{k+1}\cdots i_n}|i_{k+1}\cdots i_n\rangle)_{j_{k+1}\cdots j_n},
 \end{array}
\end{equation}
then
\begin{equation}\label{}
 \rho=|\psi\rangle\langle\psi|
 =\sum\limits_{i_1,i_2,\cdots, i_n\atop i'_1,i'_2,\cdots, i'_n}a_{i_1i_2\cdots i_k}b_{i_{k+1}\cdots i_n}a^*_{i'_1i'_2\cdots i'_k}b^*_{i'_{k+1}\cdots i'_n}
 |i_1i_2\cdots
  i_n\rangle_{j_1j_2\cdots j_n}\langle i'_1i'_2\cdots i'_n|,
\end{equation}
where
   $\{j_1,j_2,\cdots, j_n \}=\{1,2,\cdots,
n\}$. From
\begin{equation}\label{}
\begin{array}{c}
 \rho_{1,2^n} = a_{00\cdots 0}b_{00\cdots 0}a^*_{11\cdots 1}b^*_{11\cdots 1}, \\
   \rho_{\sum\limits_{l=1}^k 2^{n-j_l}+1,\sum\limits_{l=1}^k 2^{n-j_l}+1}= |a_{11\cdots 1}b_{00\cdots 0}|^2, \\
   \rho_{\sum\limits_{l=k+1}^n 2^{n-j_l}+1,\sum\limits_{l=k+1}^n 2^{n-j_l}+1} = |a_{00\cdots 0}b_{11\cdots
   1}|^2,
\end{array}
\end{equation}
 one has
\begin{equation}\label{}
   |\rho_{1,2^n}|=\sqrt{ \rho_{\sum\limits_{l=1}^k 2^{n-j_l}+1,\sum\limits_{l=1}^k 2^{n-j_l}+1}
   \rho_{\sum\limits_{l=k+1}^n 2^{n-j_l}+1,\sum\limits_{l=k+1}^n
   2^{n-j_l}+1}}.
\end{equation}
Clearly, $\sum_{l=1}^k 2^{n-j_l}+1=2,3,\cdots, 2^n-1$ for
$\{j_1,j_2,\cdots, j_n \}=\{1,2,\cdots, n\}$. Thus,
(\ref{n-qubit-biseparable}) holds for pure state $\rho$.

Next we prove that the inequality (\ref{n-qubit-biseparable}) is
also right for mixed states.

Suppose that
\begin{equation}\label{}
 \rho=\sum \limits_ip_i\rho^{(i)}=\sum
\limits_ip_i|\psi_i\rangle\langle\psi_i|
\end{equation}
is biseparable $n$-qubit state, where
$\rho^{(i)}=|\psi_i\rangle\langle\psi_i|$ is biseparable. Simple
algebra and the Cauchy inequality
$(\sum\limits_{k=1}^mx_ky_k)^2\leq(\sum\limits_{k=1}^mx_k^2)(\sum\limits_{k=1}^my_k^2)$
show that
\begin{equation}\label{}
    \begin{array}{rl}
      |\rho_{1,2^n}|= & |\sum \limits_ip_i\rho^{(i)}_{1,2^n}|\leq\sum\limits_ip_i|\rho^{(i)}_{1,2^n}| \\
     \leq  & \sum\limits_ip_i
\sum\limits_{j=2}^{2^{n-1}}\sqrt{\rho^{(i)}_{j,j}\rho^{(i)}_{2^n-j+1,2^n-j+1}}
\\
 \leq  & \sum\limits_{j=2}^{2^{n-1}} \sqrt{(\sum\limits_ip_i\rho^{(i)}_{j,j})(\sum\limits_ip_i\rho^{(i)}_{2^n-j+1,2^n-j+1})}
\\
= &
\sum\limits_{j=2}^{2^{n-1}}\sqrt{\rho_{j,j}\rho_{2^n-j+1,2^n-j+1}}.
   \end{array}
\end{equation}
The proof is complete.

The same result in this theorem has also been derived in
\cite{GuhneNJP2010}.  G\"{u}hne and Seevinck \cite{GuhneNJP2010}
proved the cases of $n=3,4$.   Here starting from general
bipartition for $n$-qubit pure states and applying the Cauchy
inequality, we give a proof for any $n$-qubit states.

Moreover, for $n$-partite and high dimension system, we have:

\textbf{Theorem 2} ~ Suppose that  $n$-partite density matrix
$\rho\in \mathcal{H}_1\otimes \mathcal{H}_2\otimes\cdots
\mathcal{H}_n$, dim$\mathcal{H}_l=d_l$, $l=1,2,\cdots, n$. If $\rho$
is biseparable, then
\begin{equation}\label{n-partite-biseparable}
 |\rho_{1, d_1d_2\cdots d_n}|\leq \frac{1}{2}\sum\limits_{i\in A}\sqrt{\rho_{i,i}\rho_{d_1d_2\cdots d_n-i+1,d_1d_2\cdots
 d_n-i+1}},
\end{equation}
where $A=\{\sum_{l=1}^{n-1}i_ld_{l+1}\cdots d_n+i_n+1 ~ | ~ i_l=0,
d_l-1, (i_1,i_2,\cdots, i_n)\neq
(0,0,\cdots,0),(d_1-1,d_2-1,\cdots,d_n-1)\}$.
 Of course, $\rho$ is a genuine $n$-partite entangled state if
it violates the above inequality (\ref{n-partite-biseparable}).

\textbf{Proof.} Suppose that $\rho=|\psi\rangle\langle\psi|$ is a
 biseparable pure state under the $j_1j_2\cdots j_k|j_{k+1}\cdots
j_n$ partition, and
\begin{equation}\label{}
 \begin{array}{rl}
   |\psi\rangle= & |\psi_1\rangle_{j_1j_2\cdots j_k}|\psi_2\rangle_{j_{k+1}\cdots j_n} \\
    = & (\sum\limits_{i_1,i_2,\cdots, i_k}a_{i_1i_2\cdots i_k}|i_1i_2\cdots i_k\rangle)_{j_1j_2\cdots j_k}
   (\sum\limits_{i_{k+1},\cdots, i_n}b_{i_{k+1}\cdots i_n}|i_{k+1}\cdots i_n\rangle)_{j_{k+1}\cdots j_n} \\
  = & \sum\limits_{i_1,i_2,\cdots, i_n}a_{i_1i_2\cdots i_k}b_{i_{k+1}\cdots i_n}|i_1i_2\cdots
  i_n\rangle_{j_1j_2\cdots j_n},
 \end{array}
\end{equation}
then
\begin{equation}\label{}
\rho_{\sum\limits_{l=1}^{n}i_ld_{j_l+1}d_{j_l+2}\cdots
d_nd_{n+1}+1,\sum\limits_{l=1}^{n}i'_ld_{j_l+1}d_{j_l+2}\cdots
d_nd_{n+1}+1}=a_{i_1i_2\cdots i_k}b_{i_{k+1}\cdots
i_n}a^*_{i'_1i'_2\cdots i'_k}b^*_{i'_{k+1}\cdots i'_n}.
\end{equation}
Here the sum is over all possible values of $i_1,i_2,\cdots, i_n$,
i.e., $\sum_{i_1,i_2,\cdots,
i_n}=\sum_{i_1=0}^{d_{j_1}-1}\sum_{i_2=0}^{d_{j_2}-1}\cdots\sum_{i_n=0}^{d_{j_n}-1}$,
$d_{n+1}=1$,
 and  $\{j_1,j_2,\cdots, j_n \}=\{1,2,\cdots,
n\}$.

  Since
\begin{equation}\label{}
\begin{array}{c}
 \rho_{1,d_1d_2\cdots d_n} = a_{00\cdots 0}b_{00\cdots 0}a^*_{d_{j_1}-1d_{j_2}-1\cdots d_{j_k}-1}b^*_{d_{j_{k+1}}-1d_{j_{k+2}}-1\cdots d_{j_n}-1}, \\
   \rho_{\sum\limits_{l=1}^k (d_{j_l}-1)d_{{j_l}+1}d_{{j_l}+2}\cdots d_nd_{n+1}+1,\sum\limits_{l=1}^k (d_{j_l}-1)d_{{j_l}+1}d_{{j_l}+2}\cdots d_nd_{n+1}+1}
   = |a_{d_{j_1}-1d_{j_2}-1\cdots d_{j_k}-1}b_{00\cdots 0}|^2, \\
   \rho_{\sum\limits_{l=k+1}^{n} (d_{j_l}-1)d_{{j_l}+1}d_{{j_l}+2}\cdots d_nd_{n+1}+1,\sum\limits_{l=k+1}^{n}(d_{j_l}-1)d_{{j_l}+1}d_{{j_l}+2}\cdots d_nd_{n+1}+1}
    = |a_{00\cdots 0}b_{d_{j_{k+1}}-1d_{j_{k+2}}-1\cdots d_{j_n}-1}|^2,
\end{array}
\end{equation}
these give
\begin{equation}\label{}
\begin{array}{rl}
  |\rho_{1,d_1d_2\cdots d_n}|= &  \sqrt{\rho_{\sum\limits_{l=1}^k (d_{j_l}-1)d_{{j_l}+1}d_{{j_l}+2}\cdots d_nd_{n+1}+1,\sum\limits_{l=1}^k (d_{j_l}-1)d_{{j_l}+1}d_{{j_l}+2}\cdots d_nd_{n+1}+1}
} \\
   & \times ~ \sqrt{\rho_{\sum\limits_{l=k+1}^{n}
(d_{j_l}-1)d_{{j_l}+1}d_{{j_l}+2}\cdots
d_nd_{n+1}+1,\sum\limits_{l=k+1}^{n}(d_{j_l}-1)d_{{j_l}+1}d_{{j_l}+2}\cdots
d_nd_{n+1}+1}}.
\end{array}
\end{equation}
 Thus, (\ref{n-partite-biseparable})
holds for pure state $\rho$.

Next we prove that the inequality (\ref{n-partite-biseparable}) is
also right for mixed states.

Suppose that
\begin{equation}\label{}
 \rho=\sum \limits_ip_i\rho^{(i)}=\sum
\limits_ip_i|\psi_i\rangle\langle\psi_i|
\end{equation}
is a biseparable $n$-partite mixed state, where
$\rho^{(i)}=|\psi_i\rangle\langle\psi_i|$ is biseparable. With the
help of (\ref{n-partite-biseparable}) for pure states $\rho^{(i)}$
and  the Cauchy inequality
$(\sum\limits_{k=1}^mx_ky_k)^2\leq(\sum\limits_{k=1}^mx_k^2)(\sum\limits_{k=1}^my_k^2)$,
there is
\begin{equation}\label{}
    \begin{array}{rl}
      |\rho_{1,d_1d_2\cdots d_n}|= & |\sum \limits_ip_i\rho^{(i)}_{1,d_1d_2\cdots d_n}|\leq\sum\limits_ip_i|\rho^{(i)}_{1,d_1d_2\cdots d_n}| \\
     \leq  &  \sum\limits_i p_i\left(\frac{1}{2}\sum\limits_{j\in A}\sqrt{\rho^{(i)}_{j,j}\rho^{(i)}_{d_1d_2\cdots d_n-j+1,d_1d_2\cdots
 d_n-j+1}}\right)  \\
  \leq  & \frac{1}{2}\sum\limits_{j\in A} \sqrt{(\sum \limits_i p_i\rho^{(i)}_{j,j}) (\sum \limits_i p_i
\rho^{(i)}_{d_1d_2\cdots d_n-j+1,d_1d_2\cdots
 d_n-j+1})}
 \\
= &
 \frac{1}{2}\sum\limits_{j\in A} \sqrt{\rho_{j,j}\rho_{d_1d_2\cdots d_n-j+1,d_1d_2\cdots
 d_n-j+1}},
  \end{array}
\end{equation}
as required.

Ineqs. (\ref{n-qubit-biseparable}) and (\ref{n-partite-biseparable})
can also be obtained from inequality (II) in
Ref.\cite{MarcusPRL2010} when $|\Phi\rangle=|00\cdots
0\rangle|11\cdots 1\rangle$ and $|\Phi\rangle=|00\cdots
0\rangle|(d_1-1)(d_2-1)\cdots (d_n-1)\rangle$, respectively.  Here
we give different proofs.

For $n$-qubit states, there is:

\textbf{Theorem 3} ~ Let $\rho$ be an  $n$-qubit state. If $\rho$ is
 biseparable, then its matrix entries fulfill
\begin{equation}\label{n-qubit-biseparable-1}
 \sum\limits_{0\leq i<j\leq n-1}|\rho_{2^i+1,2^j+1}|\leq \sum\limits_{0\leq i<j\leq n-1}\sqrt{\rho_{1,1}\rho_{2^i+2^j+1,2^i+2^j+1}}
 + \frac{n-2}{2}\sum\limits_{i=0}^{n-1}\rho_{2^i+1,2^i+1},
\end{equation}
i.e.,
\begin{equation}\label{n-qubit-biseparable-2}
 \sum\limits_{1\leq j<i\leq n}|\rho_{2^{n-i}+1,2^{n-j}+1}|\leq \sum\limits_{1\leq j<i\leq n}\sqrt{\rho_{1,1}\rho_{2^{n-i}+2^{n-j}+1,2^{n-i}+2^{n-j}+1}}
 + \frac{n-2}{2}\sum\limits_{i=1}^{n}\rho_{2^{n-i}+1,2^{n-i}+1}.
\end{equation}
If $n$-qubit state $\rho$ does not satisfy the above inequality
(\ref{n-qubit-biseparable-1}) or (\ref{n-qubit-biseparable-2}), then
$\rho$ is genuine $n$-partite entangled.

\textbf{Proof.} ~ We begin with pure state. Suppose that
$\rho=|\psi\rangle\langle\psi|$, where
$|\psi\rangle=|\phi_1\rangle_{m_1m_2\cdots
m_k}|\phi_2\rangle_{m_{k+1}\cdots m_n}$, $\{m_1,m_2,\cdots,
m_n\}=\{1,2,\cdots, n\}$. For any $1\leq j<i\leq n$, it is not
difficult to prove that
\begin{equation}\label{n-q-b-pure-1}
\begin{array}{rl}
  |\rho_{2^{n-i}+1,2^{n-j}+1}| & =\sqrt{\rho_{2^{n-i}+1,2^{n-i}+1}\rho_{2^{n-j}+1,2^{n-j}+1}} \\
   & \leq \frac{\rho_{2^{n-i}+1,2^{n-i}+1}+\rho_{2^{n-j}+1,2^{n-j}+1}}{2}
\end{array}
\end{equation}
in the case either $i,j\in A$ or $i,j\in B$, and
\begin{equation}\label{n-q-b-pure-2}
\begin{array}{c}
  |\rho_{2^{n-i}+1,2^{n-j}+1}|  = \sqrt{\rho_{1,1}\rho_{2^{n-i}+2^{n-j}+1,2^{n-i}+2^{n-j}+1}}
\end{array}
\end{equation}
in the case one of $i$ and $j$  in $A$ while another in $B$ (either
$i\in A$, $j\in B$, or $i\in B$, $j\in A$). Here
$A=\{m_1,m_2,\cdots, m_k\}$ and $B=\{m_{k+1},m_{k+2}\cdots, m_n\}$.
Combining (\ref{n-q-b-pure-1}) and (\ref{n-q-b-pure-2}) gives that
\begin{equation}\label{}
 \begin{array}{rl}
  & \sum\limits_{1\leqslant j<i\leqslant n}|\rho_{2^{n-i}+1,2^{n-j}+1}|   \\
   = & \sum\limits_{ 1\leqslant j<i\leqslant n \atop i\in A,j\in B} |\rho_{2^{n-i}+1,2^{n-j}+1}|
   +\sum\limits_{ 1\leqslant j<i\leqslant n \atop j\in A,i\in B }|\rho_{2^{n-i}+1,2^{n-j}+1}|
   + \sum\limits_{1\leqslant j<i\leqslant n \atop i,j\in A }|\rho_{2^{n-i}+1,2^{n-j}+1}|
 +\sum\limits_{1\leqslant j<i\leqslant n \atop i,j\in B }|\rho_{2^{n-i}+1,2^{n-j}+1}|   \\
 \leq & \sum\limits_{ 1\leqslant j<i\leqslant n \atop i\in A,j\in B}
 \sqrt{\rho_{1,1}\rho_{2^{n-i}+2^{n-j}+1,2^{n-i}+2^{n-j}+1}}+
  \sum\limits_{ 1\leqslant j<i\leqslant n \atop j\in A,i\in B}
  \sqrt{\rho_{1,1}\rho_{2^{n-i}+2^{n-j}+1,2^{n-i}+2^{n-j}+1}} \\
& +\sum\limits_{1\leqslant j<i\leqslant n \atop i,j\in A
}\frac{\rho_{2^{n-i}+1,2^{n-i}+1}+\rho_{2^{n-j}+1,2^{n-j}+1}}{2}
  +\sum\limits_{1\leqslant j<i\leqslant n \atop i,j\in B }\frac{\rho_{2^{n-i}+1,2^{n-i}+1}+\rho_{2^{n-j}+1,2^{n-j}+1}}{2}  \\
\leq & \sum\limits_{1\leq j<i\leq
n}\sqrt{\rho_{1,1}\rho_{2^{n-i}+2^{n-j}+1,2^{n-i}+2^{n-j}+1}}
 + \frac{n-2}{2}\sum\limits_{i=1}^{n}\rho_{2^{n-i}+1,2^{n-i}+1}.
  \end{array}
\end{equation}
that is, (\ref{n-qubit-biseparable-2}) holds for any biseparable
$n$-qubit pure state $\rho$.

Now we suppose that $\rho=\sum \limits_mp_m\rho^{(m)}$ is a
biseparable  mixed state, and
$\rho^{(m)}=|\psi_m\rangle\langle\psi_m|$ is biseparable. Then,
simple algebra and the Cauchy inequality show that

\begin{equation}\label{}
\begin{array}{rl}
   &  \sum\limits_{1\leqslant j<i\leqslant n}|\rho_{2^{n-i}+1,2^{n-j}+1}|   \\
   = & \sum\limits_{1\leqslant j<i\leqslant n}|\sum\limits_{m}p_m
   \rho^{(m)}_{2^{n-i}+1,2^{n-j}+1}| \\
 \leq & \sum\limits_{m}p_m\sum\limits_{1\leqslant j<i\leqslant n}
  |\rho^{(m)}_{2^{n-i}+1,2^{n-j}+1}| \\
\leq &   \sum\limits_{m}p_m \left(\sum\limits_{1\leq j<i\leq
n}\sqrt{\rho^{(m)}_{1,1}\rho^{(m)}_{2^{n-i}+2^{n-j}+1,2^{n-i}+2^{n-j}+1}}
 +
 \frac{n-2}{2}\sum\limits_{i=1}^{n}\rho^{(m)}_{2^{n-i}+1,2^{n-i}+1}\right)
 \\
= & \sum\limits_{1\leq j<i\leq n}\sum\limits_{m}
\sqrt{p_m\rho^{(m)}_{1,1}}\sqrt{p_m\rho^{(m)}_{2^{n-i}+2^{n-j}+1,2^{n-i}+2^{n-j}+1}}
 +
 \frac{n-2}{2}\sum\limits_{i=1}^{n}\sum\limits_{m}p_m
 \rho^{(m)}_{2^{n-i}+1,2^{n-i}+1}\\
\leq & \sum\limits_{1\leq j<i\leq n}
\sqrt{\sum\limits_{m}p_m\rho^{(m)}_{1,1}}\sqrt{\sum\limits_{m}p_m\rho^{(m)}_{2^{n-i}+2^{n-j}+1,2^{n-i}+2^{n-j}+1}}
 +
 \frac{n-2}{2}\sum\limits_{i=1}^{n}\sum\limits_{m}p_m
 \rho^{(m)}_{2^{n-i}+1,2^{n-i}+1} \\
= &  \sum\limits_{1\leq j<i\leq
n}\sqrt{\rho_{1,1}\rho_{2^{n-i}+2^{n-j}+1,2^{n-i}+2^{n-j}+1}}
 + \frac{n-2}{2}\sum\limits_{i=1}^{n}\rho_{2^{n-i}+1,2^{n-i}+1},
\end{array}
\end{equation}
which is the desired result.

Observation 3 and Observation 4 (ii) in \cite{GuhneNJP2010} are the
special cases $n=3$ and $n=4$ of Theorem 3, respectively.

\section{The separability criteria of fully separable $n$-partite states }

 In this section, we consider fully separable $n$-partite states.

 For fully separable $n$-qubit states, by utilizing the Cauchy inequality and H\"{o}lder inequality,  we derive:

\textbf{Theorem 4} ~ If an $n$-qubit density matrix $\rho$ is fully
separable, then the following inequalities hold:
\begin{equation}\label{n-qubit-fully-separable}
 |\rho_{1,2^n}|\leq
 \left(\rho_{2,2}\rho_{3,3}\rho_{4,4}\cdots\rho_{2^n-1,2^n-1}\right)^{\frac{1}{2^n-2}},
\end{equation}
\begin{equation}\label{n-qubit-fully-separable-1}
\sum\limits_{0\leq i<j\leq n-1}|\rho_{2^i+1,2^j+1}|\leq
\sum\limits_{0\leq i<j\leq
n-1}\sqrt{\rho_{1,1}\rho_{2^i+2^j+1,2^i+2^j+1}}.
\end{equation}
These two inequalities are equalities for fully separable
$n$-partite pure states.

\textbf{Proof.} First, let us start with pure states.

Suppose that $\rho=|\psi\rangle\langle\psi|$ is a fully sepaprable
$n$-qubit pure state, where
\begin{equation}\label{}
\begin{array}{rl}
 |\psi\rangle&=\left(a_{10}|0\rangle+a_{11}|1\rangle\right)\otimes\left(a_{20}|0\rangle+a_{21}|1\rangle\right)\otimes
 \cdots\otimes\left(a_{n0}|0\rangle+a_{n1}|1\rangle\right)\\
&=\sum\limits_{i_1,\cdots,i_n=0}^1a_{1i_1}a_{2i_2}\cdots
a_{ni_n}|i_1i_2\cdots i_n\rangle.
\end{array}
\end{equation}
Then
\begin{equation}\label{}
 \rho_{i,j}=a_{1i_1}a_{2i_2}\cdots
a_{ni_n} a_{1j_1}^*a_{2j_2}^*\cdots a_{nj_n}^*,
\end{equation}
where $i=\sum_{k=1}^ni_k\cdot2^{n-k}+1,
j=\sum_{k=1}^nj_k\cdot2^{n-k}+1$. It follows that
\begin{equation}\label{}
  \begin{array}{rl}
     & \rho_{2,2}\rho_{3,3}\cdots\rho_{2^n-1,2^n-1} \\
   = & |a_{10}a_{20}\cdots
a_{n-10}a_{n1}|^2|a_{10}a_{20}\cdots
a_{n-11}a_{n0}|^2\cdots|a_{11}a_{21}\cdots
a_{n-11}a_{n0}|^2\\
=&|a_{10}a_{20}\cdots a_{n0}a_{11}a_{21}\cdots a_{n1}|^{2^n-2} \\
=& (\rho_{1,2^n})^{2^n-2},
  \end{array}
\end{equation}
that is, the inequality (\ref{n-qubit-fully-separable}) is an
equality for fully separable $n$-qubit pure states.

Note that
\begin{equation}\label{}
\begin{array}{cl}
   & \rho_{\sum_{l=1}^t 2^{n-k_l}+1,\sum_{l=1}^t 2^{n-k_l}+1}\rho_{\sum_{l=t+1}^n 2^{n-k_l}+1,\sum_{l=t+1}^n 2^{n-k_l}+1} \\
 = & |a_{k_11}a_{k_21}\cdots a_{k_t1}a_{k_{t+1}0}\cdots a_{t_n0}|^2|a_{k_10}a_{k_20}\cdots a_{k_t0}a_{k_{t+1}1}\cdots
 a_{t_n1}|^2 \\
 = & |a_{10}a_{20}\cdots a_{n0}a_{11}a_{21}\cdots a_{n1}|^2
 \\
 = & |\rho_{1,2^n}|^2
\end{array}
\end{equation}
for any $\{k_1,k_2,\cdots,k_n\}=\{1,2,\cdots,n\}$.
 It also
implies that the inequality (\ref{n-qubit-fully-separable}) is an
equality for fully separable $n$-qubit pure states.

(\ref{n-qubit-fully-separable-1}) follows immediately from
\begin{equation}\label{}
|\rho_{2^i+1,2^j+1}|=\sqrt{\rho_{1,1}\rho_{2^i+2^j+1,2^i+2^j+1}}.
\end{equation}

Next we show that the inequality (\ref{n-qubit-fully-separable}) is
also right for fully separable mixed states.

Suppose that $\rho=\sum \limits_ip_i\rho^{(i)}$, where  $\rho^{(i)}$
is fully separable $n$-qubit pure state. Then
\begin{equation}\label{}
|\rho_{1,2^n}|=|\sum \limits_ip_i\rho_{1,2^n}^{(i)}|\leq\sum
\limits_ip_i|\rho_{1,2^n}^{(i)}|=\sum
\limits_ip_i(\rho_{2,2}^{(i)}\rho_{3,3}^{(i)}\cdots\rho_{2^n-1,2^n-1}^{(i)})^{\frac{1}{2^n-2}}.
\end{equation}
Continuously using the H\"{o}lder inequality
\begin{equation}\label{}
\sum\limits_{k=1}^m|x_ky_k|\leq(\sum\limits_{k=1}^m|x_k|^p)^{\frac{1}{p}}(\sum\limits_{k=1}^m|y_k|^q)^{\frac{1}{q}}
~ (p,q>1,\dfrac{1}{p}+\dfrac{1}{q}=1),
\end{equation}
 we get
\begin{equation}\label{}
\begin{array}{rl}
&  \sum
\limits_ip_i(\rho_{2,2}^{(i)}\rho_{3,3}^{(i)}\cdots\rho_{2^n-1,2^n-1}^{(i)})^{\frac{1}{2^n-2}}
\\
 = & \sum
\limits_i(p_i\rho_{2,2}^{(i)})^\frac{1}{2^n-2}(p_i\rho_{3,3}^{(i)}\cdots
p_i\rho_{2^n-1,2^n-1}^{(i)})^{\frac{1}{2^n-2}} \\
\leq &\left(\sum \limits_i
p_i\rho_{2,2}^{(i)}\right)^\frac{1}{2^n-2}\left[\sum
\limits_i(p_i\rho_{3,3}^{(i)}\cdots
p_i\rho_{2^n-1,2^n-1}^{(i)})^{\frac{1}{2^n-3}}\right]^{\frac{2^n-3}{2^n-2}} \\
\leq & \left(\sum \limits_i
p_i\rho_{2,2}^{(i)}\right)^\frac{1}{2^n-2} \left[\left(\sum
\limits_i p_i\rho_{3,3}^{(i)}\right)^\frac{1}{2^n-3}\left(\sum
\limits_i(~p_i\rho_{4,4}^{(i)}\cdots
p_i\rho_{2^n-1,2^n-1}^{(i)})^{\frac{1}{2^n-4}}\right)^{\frac{2^n-4}{2^n-3}}\right]^{\frac{2^n-3}{2^n-2}}
\\
= &  \left(\sum \limits_i p_i\rho_{2,2}^{(i)}\right)^\frac{1}{2^n-2}
\left(\sum \limits_i
p_i\rho_{3,3}^{(i)}\right)^\frac{1}{2^n-2}\left[\sum
\limits_i(~p_i\rho_{4,4}^{(i)}\cdots
p_i\rho_{2^n-1,2^n-1}^{(i)})^{\frac{1}{2^n-4}}\right]^{\frac{2^n-4}{2^n-2}}
\\
\leq & \left(\sum \limits_i
p_i\rho_{2,2}^{(i)}\right)^\frac{1}{2^n-2} \left(\sum \limits_i
p_i\rho_{3,3}^{(i)}\right)^\frac{1}{2^n-2}\cdots\left(\sum \limits_i
p_i\rho_{2^n-1,2^n-1}^{(i)}\right)^\frac{1}{2^n-2} \\
= &
(\rho_{2,2}\rho_{3,3}\cdots\rho_{2^n-1,2^n-1})^{\frac{1}{2^n-2}},
\end{array}
\end{equation}
as claimed.

Simple algebra and the Cauchy inequality show that
(\ref{n-qubit-fully-separable-1}) holds for fully separable
$n$-partite mixed states.

Observation 4 (i) and (iii) in \cite{GuhneNJP2010} are the case
$n=3$ of this theorem.

For the well-studied $n$-qubit GHZ states mixed with white noise,
Theorem 4 constitutes a necessary and sufficient criterion for fully
separable.

\textbf{Theorem 5} ~ For
$\rho(p)=(1-p)|\textrm{GHZ}_n\rangle\langle\textrm{GHZ}_n|+\dfrac{p}{2^n}\textrm{I}$,
 $\rho(p)$ is fully separable iff the entries of $\rho(p)$ satisfy
 the
inequality (\ref{n-qubit-fully-separable}).

\textbf{Proof.} Necessity is immediate from Theorem 4. Conversely if
the inequality (\ref{n-qubit-fully-separable}) holds for $\rho(p)$,
i.e.
$|\rho(p)_{1,2^n}|\leq\left(\rho(p)_{2,2}\rho(p)_{3,3}\rho(p)_{4,4}\cdots\rho(p)_{2^n-1,2^n-1}\right)^{\frac{1}{2^n-2}}$,
then there is
$\frac{1-p}{2}\leq\left[(\frac{p}{2^n})^{2^n-2}\right]^\frac{1}{2^n-2}$,
which implies that $p\geq1-\frac{1}{2^{n-1}+1}$. Therefore,
$\rho(p)$ is fully separable \cite{DurCiracPRA99}.

Observation 4 (iv) in \cite{GuhneNJP2010} is the case $n=3$ of this
theorem.

Furthermore, for high dimension and $n$-partite, using the
H\"{o}lder inequality, we can infer:

\textbf{Theorem 6} ~ For any $n$-particle density matrix $\rho$
(particle $k$ is $d_k$ level, $ 1\leq k\leq n$), if $\rho$ is fully
separable, then
\begin{equation}\label{n-partite-fully-separable}
|\rho_{1,d_1d_2\cdots d_n}|\leq (\prod\limits_{i\in
A}\rho_{ii})^{\frac{1}{2^n-2}},
\end{equation}
where $A$ is the set of $2^n-2$ numbers
$\sum_{k=1}^{n-1}i_kd_{k+1}d_{k+2}\cdots d_n+i_n+1$ such that
$i_k\in \{0,d_k-1\}$, and $(i_1,i_2,\cdots,
i_n)\neq(0,0,\cdots,0),(d_1-1,d_2-1,\cdots, d_n-1)$, i.e.,
   $A=\{i=\sum_{k=1}^{n-1}i_kd_{k+1}d_{k+2}\cdots
d_n+i_n+1~|~ i_k=0,d_k-1, ~ k=1,2,\cdots,n, ~ i\neq 1, i\neq
d_1d_2\cdots d_n\}$.

If $\rho$ is a fully separable $n$-particle pure state, then the
inequality (\ref{n-partite-fully-separable}) is an equality.

\textbf{Proof.} ~ Suppose that $\rho=|\psi\rangle\langle\psi|$ is
fully separable pure state, where
\begin{equation}\label{}
 \begin{array}{rl}
   |\psi\rangle= & (\sum\limits_{i_1=0}^{d_1-1}a_{1i_1}|i_1\rangle)\otimes(\sum\limits_{i_2=0}^{d_2-1}a_{1i_2}|i_2\rangle)\otimes
   \cdots\otimes(\sum\limits_{i_n=0}^{d_n-1}a_{1i_n}|i_n\rangle) \\
   = & \sum\limits_{i_1=0}^{d_1-1}\sum\limits_{i_2=0}^{d_2-1}\cdots\sum\limits_{i_n=0}^{d_n-1}a_{1i_1}a_{2i_2}\cdots
a_{ni_n}|i_1i_2\cdots i_n\rangle.
 \end{array}
\end{equation}
Then the elements of $\rho$
\begin{equation}\label{}
\rho_{i,j}=a_{1i_1}a_{2i_2}\cdots a_{ni_n}a^*_{1j_1}a^*_{2j_2}\cdots
a^*_{nj_n},
\end{equation}
where $i=\sum_{k=1}^{n-1}i_kd_{k+1}d_{k+2}\cdots d_n+i_n+1$,
$j=\sum_{k=1}^{n-1}j_kd_{k+1}d_{k+2}\cdots d_n+j_n+1$.

 Since
\begin{equation}\label{}
 \begin{array}{rl}
   & \rho_{\sum\limits_{l=1}^{t}(d_{k_l}-1)d_{k_l+1}\cdots d_nd_{n+1}+1,\sum\limits_{l=1}^{t}(d_{k_l}-1)d_{k_l+1}\cdots d_nd_{n+1}+1}
   \rho_{\sum\limits_{l=t+1}^{n}(d_{k_l}-1)d_{k_l+1}\cdots d_nd_{n+1}+1,\sum\limits_{l=t+1}^{n}(d_{k_l}-1)d_{k_l+1}\cdots d_nd_{n+1}+1} \\
   = & |a_{k_1d_{k_1}-1}a_{k_2d_{k_2}-1}\cdots a_{k_td_{k_t}-1}a_{k_{t+1}0}\cdots
   a_{k_n0}|^2 |a_{k_10}a_{k_20}\cdots a_{k_t0}a_{k_{t+1}d_{k_{t+1}}-1}\cdots
   a_{k_nd_{k_n}-1}|^2 \\\
   =& |a_{10}a_{20}\cdots a_{n0}a_{1d_1-1}a_{2d_2-1}\cdots
   a_{nd_n-1}|^2 \\
   = & |\rho_{1,d_1d_2\cdots d_n}|^2,
 \end{array}
\end{equation}
for any $\{k_1,k_2,\cdots,k_t,k_{t+1},\cdots,
k_n\}=\{1,2,\cdots,n\}$ and $d_{n+1}=1$, this gives
\begin{equation}\label{}
\begin{array}{rl}
& \left(|\rho_{1,d_1d_2\cdots d_n}|^2\right)^{2^n-2} \\
 = & \prod\limits_{\{k_1,\cdots,k_t,k_{t+1},\cdots,
k_n\}\atop
=\{1,2,\cdots,n\}}\rho_{\sum\limits_{l=1}^{t}(d_{k_l}-1)d_{k_l+1}\cdots
d_nd_{n+1}+1,\sum\limits_{l=1}^{t}(d_{k_l}-1)d_{k_l+1}\cdots
d_nd_{n+1}+1}
   \rho_{\sum\limits_{l=t+1}^{n}(d_{k_l}-1)d_{k_l+1}\cdots d_nd_{n+1}+1,\sum\limits_{l=t+1}^{n}(d_{k_l}-1)d_{k_l+1}\cdots d_nd_{n+1}+1} \\
 =  & (\prod\limits_{i\in A}\rho_{ii})^2.
\end{array}
\end{equation}
It implies that
\begin{equation}\label{n-partite-fully-separable-pure}
|\rho_{1,d_1d_2\cdots d_n}|=(\prod\limits_{i\in
A}\rho_{ii})^{\frac{1}{2^n-2}},
\end{equation}
thus (\ref{n-partite-fully-separable}) holds for  fully separable
pure states. Here  $A=\{i=\sum_{k=1}^{n-1}i_kd_{k+1}d_{k+2}\cdots
d_n+i_n+1~|~ i_k=0,d_k-1, ~ k=1,2,\cdots,n, ~ i\neq 1, i\neq
d_1d_2\cdots d_n\}$.

One can also derive (\ref{n-partite-fully-separable-pure}) by direct
calculation.

Next we suppose that $\rho=\sum_ip_i\rho^{(i)}$ is an $n$-partite
mixed state, where $\rho^{(i)}=|\psi^i\rangle\langle\psi^i|$ is
fully separable. Using (\ref{n-partite-fully-separable-pure}) for
each $\rho^{(i)}$, we see
\begin{equation}\label{n-partite-full-separable-1}
\begin{array}{rl}
  |\rho_{1,d_1d_2\cdots d_n}|& =|\sum \limits_ip_i\rho_{1,d_1d_2\cdots d_n}^{(i)}| \\
  &   \leq\sum
\limits_ip_i|\rho_{1,d_1d_2\cdots d_n}^{(i)}|=\sum
\limits_ip_i(\prod\limits_{j\in{A}}\rho_{jj}^{(i)})^{\frac{1}{2^n-2}}.
\end{array}
\end{equation}
Let $m_2, m_3,\cdots, m_{2^n-1}$  be the elements in the set $A$. By
the H\"{o}lder inequality, we obtain
\begin{equation}\label{n-partite-full-separable-2}
\begin{array}{rl}
   & \sum
\limits_ip_i(\prod\limits_{j\in{A}}\rho_{jj}^{(i)})^{\frac{1}{2^n-2}} \\
 = & \sum
\limits_i(p_i\rho_{m_2,m_2}^{(i)})^\frac{1}{2^n-2}(p_i\rho_{m_3,m_3}^{(i)}\cdots
p_i\rho_{m_{2^n-1},m_{2^n-1}}^{(i)})^{\frac{1}{2^n-2}}\\
\leq & \left(\sum \limits_i
p_i\rho_{m_2,m_2}^{(i)}\right)^\frac{1}{2^n-2}\left[\sum
\limits_i(p_i\rho_{m_3,m_3}^{(i)}\cdots
p_i\rho_{m_{2^n-1},m_{2^n-1}}^{(i)})^{\frac{1}{2^n-3}}\right]^{\frac{2^n-3}{2^n-2}} \\
\leq & \left(\sum \limits_i
p_i\rho_{m_2,m_2}^{(i)}\right)^\frac{1}{2^n-2} \left(\sum \limits_i
p_i\rho_{m_3,m_3}^{(i)}\right)^\frac{1}{2^n-2} \left[\sum
\limits_i(p_i\rho_{m_4,m_4}^{(i)}\cdots
p_i\rho_{m_{2^n-1},m_{2^n-1}}^{(i)})^{\frac{1}{2^n-4}}\right]^{\frac{2^n-4}{2^n-2}}
\\
\leq &\left[\sum
\limits_i(p_i\rho_{m_2,m_2}^{(i)})\right]^\frac{1}{2^n-2}\left[\sum
\limits_i(p_i\rho_{m_3,m_3}^{(i)})\right]^\frac{1}{2^n-2}\cdots\left[\sum
\limits_i(p_i\rho_{m_{2^n-1},m_{2^n-1}}^{(i)})\right]^{\frac{1}{2^n-2}}\\
=&(\rho_{m_2,m_2}\rho_{m_3,m_3}\cdots\rho_{m_{2^n-1},m_{2^n-1}})^{\frac{1}{2^n-2}}\\
=&(\prod\limits_{i\in{A}}\rho_{ii})^{\frac{1}{2^n-2}}.
\end{array}
\end{equation}
Combining (\ref{n-partite-full-separable-1}) and
(\ref{n-partite-full-separable-2}) gives the inequality
(\ref{n-partite-fully-separable}), as required.

\section{conclusion}

We derive separability criteria for $n$-qubit and $n$-qudit quantum
states directly in terms of matrix elements. Some of them are also
sufficient conditions for   genuine entanglement of $n$-partite
quantum states.  One of the resulting criteria is also necessary and
sufficient condition for a class of $n$-partite states. We give
clear and complete proof of each criterion from general partition by
using the Cauchy inequality and H\"{o}lder inequality.
\\

This work was supported by the National Natural Science Foundation
of China under Grant No: 10971247, Hebei Natural Science Foundation
of China under Grant Nos: F2009000311, A2010000344.

\end{document}